\documentclass[aip,jcp,
amsmath,amssymb,
reprint,
]{revtex4-2}

\usepackage{graphicx}
\usepackage{dcolumn}
\usepackage{bm}

\usepackage[utf8]{inputenc}
\usepackage[T1]{fontenc}
\usepackage{mathptmx}
\usepackage{etoolbox}
\usepackage{mathrsfs}
\usepackage{color,colordvi}
\usepackage{comment}

\makeatletter
\def\@email#1#2{%
 \endgroup
 \patchcmd{\titleblock@produce}
  {\frontmatter@RRAPformat}
  {\frontmatter@RRAPformat{\produce@RRAP{*#1\href{mailto:#2}{#2}}}\frontmatter@RRAPformat}
  {}{}
}%
\makeatother
\begin{document}

\preprint{AIP/123-QED}

\title{Sampling the Liquid-Gas Critical Point with Boltzmann Generators}

\author{Luigi de Santis}
\affiliation{Department of Physics, Sapienza University of Rome, Piazzale Aldo Moro 2, 00185 Roma, Italy}
\author{John Russo}
\affiliation{Department of Physics, Sapienza University of Rome, Piazzale Aldo Moro 2, 00185 Roma, Italy}
\author{Andrea Ninarello}
 \email{andrea.ninarello@cnr.it}
\affiliation{CNR Institute of Complex Systems, Uos Sapienza, Piazzale Aldo Moro 2, 00185, Roma, Italy}
\affiliation{Department of Physics, Sapienza University of Rome, Piazzale Aldo Moro 2, 00185 Roma, Italy}
 
\date{\today}

\begin{abstract}
Generative models based on invertible transformations provide a physics-aware route to sample equilibrium configurations directly from the Boltzmann distribution, enabling efficient exploration of complex thermodynamic landscapes.  Here, we evaluate their applicability in regions where conventional simulations suffer from severe dynamical bottlenecks, focusing on the liquid–gas critical point of a Lennard–Jones fluid. We show that Boltzmann Generators capture essential signatures of critical behavior, retain reliable performance when trained at or near criticality, and extrapolate across neighboring states of the phase diagram. An intriguing observation is that the model’s efficiency metric closely traces the underlying phase boundaries, hinting at a connection between generative performance and thermodynamics. However, the approach remains limited by the small system sizes currently accessible, which suppress the large fluctuations that characterize critical phenomena. Our results delineate the current capabilities and boundaries of Boltzmann Generators in challenging regions of phase space, while pointing toward future applications in problems dominated by slow dynamics, such as glass formation and nucleation.

\end{abstract}

\maketitle

\section{Introduction}

Since the mid-20th century, molecular dynamics and Monte Carlo simulations have been primary tools for exploring the phase behavior of liquids and understanding their equilibrium thermodynamics.~\cite{Frenkel2001, Newman1998} These techniques remain foundational, as their primary objective is to reconstruct equilibrium distributions by sampling configurational ensembles, whether through stochastic MC moves or deterministic MD trajectories. While simulations excel at exploring accessible regions of phase diagrams, their utility is challenged in systems exhibiting slow equilibration dynamics, metastability, or critical phenomena. Examples include glass-forming liquids hindered by rugged energy landscapes,~\cite{berthier2023} nucleation processes requiring rare-event sampling,~\cite{ji2016} and phase transitions in liquids plagued by critical slowing down.~\cite{sciortino2000,dijkstra2001,wilson2005}

To address these challenges, enhanced sampling methods such as metadynamics~\cite{barducci2011}, umbrella sampling,~\cite{earl2005} and parallel tempering~\cite{kastner2011} have emerged that effectively circumvent kinetic barriers by biasing simulations or accelerating phase-space exploration. Yet, the growing complexity of soft matter systems and the demand for high-resolution phase diagrams have spurred interest in integrating machine learning into computational frameworks. Initially, machine learning tools were deployed to classify phases, detect transitions, or analyze order parameters.~\cite{wang2016,carrasquilla2017,mendes2021} However, recent advances have shifted their application toward replacing or enhancing traditional simulation methodologies, particularly in sampling elusive regions of phase space.~\cite{mehdi2024}

Notably, ML-driven interatomic potentials now enable accurate and efficient modeling of many-body interactions,~\cite{behler2016} while adaptive sampling techniques leverage reinforcement learning to prioritize underrepresented states.~\cite{kleiman2023} Diffusion models, inspired by non-equilibrium thermodynamics, have also shown promise in generating equilibrium configurations by iteratively denoising distributions.~\cite{yang2023} Among generative AI approaches, normalizing flows have garnered significant attention.~\cite{kobyzev2020,papamakarios2021} Normalizing flows employ sequences of invertible, learnable transformations to map simple base distributions (e.g., Gaussians) to complex target distributions. By embedding thermodynamic principles into their framework, models like Boltzmann Generators (BGs) modify standard normalizing flows to sample directly from the Boltzmann distribution.~\cite{noe2019} This is achieved by combining invertible transformations with energy-based learning, allowing the model to generate equilibrium configurations consistent with statistical mechanics. The transformation invertibility ensures exact density estimation, enabling direct sampling of equilibrium states while preserving physical interpretability. Therefore, normalizing flows are able to bridge data-driven learning and physics-informed modeling, offering a transformative framework for exploring challenging regions of phase diagrams.

Boltzmann Generators have been recently emerged as powerful tools for sampling complex, high-dimensional distributions in many-particle system models. They have been demonstrated to accelerate MC by combining local moves with learned nonlocal transitions~\cite{gabrie2022}, including in the context of rare-event sampling~\cite{asghar2024}, enable free energy estimation of atomic solids without requiring ground-truth samples,~\cite{wirnsberger2022} and facilitate unsupervised training for predicting thermodynamic properties in solids~\cite{ahmad2022}. Extensions to the isobaric-isothermal (NPT) ensemble allow direct sampling of pressure-dependent systems~\cite{wirnsberger2023}, and their applicability has been demonstrated in studies of glass-forming liquids~\cite{jung2024, jung2025}.


A recent line of work has shown that Boltzmann Generators can be conditioned on thermodynamic state variables, leading to so-called conditional normalizing flows that enable training within a specific region of a phase diagram and facilitate the generation of configurations across neighboring states~\cite{wirnsberger2020, schebek2024}. Notably, this framework builds on physically informed flow architectures that exploit permutation equivariance to efficiently model transformations between physical distributions. While this approach shows promise for exploring phase space and computing free-energy differences, it remains unclear whether it can effectively address particularly challenging regions, such as those affected by critical slowing down.

In this work, we examine this question by employing BGs in the conditional setting, as defined in Ref.~\cite{schebek2024}, to generate configurations around the liquid-gas critical point of a Lennard-Jones system. We first benchmark the method in a defined liquid phase. We then train BGs directly at the critical point and assess their ability to extrapolate to nearby states. Conversely, we also train away from the critical point and evaluate performance as the system approaches criticality. 
We validate our findings by examining both energy averages and their distributions. We further compare the expected and generated heat capacity and compressibility, as these fluctuation-based quantities are essential for characterizing critical phenomena.
Our results, building on established normalizing-flow methods and architectures, indicate that BGs achieve strong performance close to criticality while still maintaining effectiveness when extrapolated beyond the critical point. However, their efficiency is offset by the small system sizes accessible, since limited access to high-dimensional spaces suppresses large-scale fluctuations. Taken together, these findings delineate the current scope of BGs and point toward their future potential in addressing complex phenomena such as glass formation and nucleation.

\section{Methods}

\subsection{Model}
We investigate the phase behavior of a Lennard-Jones (LJ) system under isothermal-isobaric (NPT) conditions, focusing on mapping the solid-liquid coexistence line. The system consists of $N=180$ particles, initially arranged in a disordered, non-overlapping configuration. While the system size might seem limited, it is consistent with current state-of-the-art practices~\cite{schebek2024,jung2024}. 
Periodic boundary conditions are applied, and isotropic volume fluctuations are permitted to maintain constant pressure. The LJ potential
\begin{equation}
    V_{LJ} = 4\epsilon \left[\left(\frac{\sigma}{r}\right)^{12}-\left(\frac{\sigma}{r}\right)^{6}\right]
\end{equation}
is truncated at a cutoff distance $r_c=2.2\sigma$, with a multiplicative smoothing function, applied between $r_s=0.9r_c$ and $r_c$ to ensure continuity, of the form
\begin{equation}
S(x)=1-6x^5+15x^4-10x^3
\end{equation}
where $x=\frac{r-r_s}{r_c-r_s}$.
In the following, all results are reported in reduced units of length $\sigma$ and energy $\epsilon$.

\noindent In the context of normalizing flows, the prior defines the base distribution from which samples are drawn and subsequently transformed to approximate the target distribution. In our case, the prior state is represented by 20,000 equilibrium configurations obtained through conventional Metropolis Monte Carlo simulations. These configurations were generated over 40 million Monte Carlo steps, with one sample collected every 2,000 steps to ensure statistical independence. In the following, we refer to this distribution as the prior when it is used during training. However, in the conditional setting described below, it may also coincides with the reference distribution against which the network efficiency is evaluated.

\subsection{Conditional Normalizing Flows}
As anticipated, a Normalizing Flow is composed of a series of invertible and differentiable transformations, concatenated to transform a simple prior distribution into a target distribution. Interestingly, the prior and target distributions may correspond to either different or identical thermodynamic states in the NPT ensemble, thereby defining an NPT flow.
Moreover, Normalizing Flows can be extended to a conditional setting, which allows them to model conditional target distributions. A Conditional Normalizing Flow learns a mapping from a prior distribution $p_A(\mathbf{x})$ to a conditional target distribution $ p_B(\mathbf{x} \mid \mathbf{c}) $, where $ \mathbf{c} $ represents the conditioning variables. These conditional variables are incorporated into the transformation functions as additional inputs, enabling the model to generate or evaluate samples that depend on the given context or conditions. While originally developed for image generation purpose~\cite{ardizzone2019, winkler2019}, Conditional Normalizing Flows have recently found applications in the sampling of rare events~\cite{falkner2023} and in lattice field theory.~\cite{singha2023}

In general the Boltzmann distribution in the NPT ensemble can be defined as
\begin{equation}
p(\mathbf{x}, V) = \frac{1}{Z} \exp\left[-\beta \left(U(\mathbf{x}, V) + PV\right)\right]
\end{equation}
\newline
\noindent where $Z$ is the partition function in the NPT ensemble, $\mathbf{x}$ is real three-dimensional vectors and $V$ is the system volume.
Let us consider two thermodynamic states, $A$ and $B$, characterized by their respective Boltzmann distributions $p_A$ and $p_B$. The goal is to learn a bijective transformation $ T((\textbf{x},V),\boldsymbol{\theta})$, parameterized by $\boldsymbol{\theta}$, that maps configurations sampled from $p_A$  to configurations with significant statistical weight under $p_B$ 
\begin{equation}
\begin{aligned}
    p'_A(T(\mathbf{x},V)~|~\mathbf{c}) & = p_A\left((\mathbf{x},V)~|~\mathbf{c}\right)  / \left| J_{T}((\mathbf{x},V)~|~\mathbf{c}) \right| \\
\end{aligned}
\end{equation}
where prime indicates the distribution approximating the target one, i.e. $p'_A(T(\mathbf{x})) \approx p_B(\mathbf{x})$ and $J_{T}$ is the Jacobian of the trasformation.
Since the similarity between two probability distributions can be measured using the Kullback-Leibler (KL) divergence, minimizing it by optimizing the model parameters $\boldsymbol{\theta}$ allows the generated distribution to more closely approximate the target distribution.
By defining the importance weigth as
\begin{equation}
   \omega_{BA'}(\mathbf{x},V) = \frac{p_B(T(\mathbf{x},V))}{p_A'(T(\mathbf{x},V))}
\end{equation}
the forward KL divergence reads:
\begin{equation}
    D_{\text{KL}}(p'_A \| p_B) = -\int p_A(\mathbf{x},V) \log(\omega_{BA'}(\mathbf{x},V)) d\mathbf{x}dV - \Delta f_{AB}
    \label{eq:KLf}
\end{equation}
where $\Delta f_{AB}$ represents the free energy difference between the two states, a parameter independent
of training, while the first term can be taken as the loss function of a training algorithm  $\mathcal{L}_F(\boldsymbol{\theta})$.

As anticipated, NPT-flows map a broad set of states beyond the initial training point by shifting from single-point learning to regional coverage. They define a range of temperatures and pressures, creating a discrete grid of thermodynamic states. During training, the model randomly selects states from this grid, generates configurations, and computes their energies to evaluate the loss as~\cite{wirnsberger2020}:
\begin{equation}
    \mathcal{L}_F(\boldsymbol{\theta}) =  \frac{U_B+V_BP_B}{k_BT_B} - \frac{U_A+V_AP_A}{k_BT_A} - \log|J_T(\mathbf{x},V)|
\end{equation}
Iterating over different sampled temperatures, the network gradually converges toward loss-minimizing states, continuing until the loss plateaus. 
We employ a learning rate $\gamma=5 \times 10^{-5}$ chosen as an optimal balance, small enough to ensure stable convergence without overshooting, yet large enough to avoid stagnation during training.

\noindent Following generation, configurations and observables are properly reweighted using the weight defined as:
\begin{equation}
    \log(\omega_{BA}(\boldsymbol{x},V))= -\frac{U_B + V_B P_B}{k_BT_B} + \frac{U_A + V_A P_A}{k_BT_A} + \log |J_T(\boldsymbol{x},V)|
\end{equation}
The code and network setup were adapted from Ref.~\cite{schebek2024} with minor modifications. We also tested various model hyperparameter configurations and did not observe any significant differences in performance.

We note in passing that, in the conditional training considered here, the model is optimized via a training by energy procedure~\cite{noe2019}. Specifically, configurations sampled from a prior distribution at $(T_0^*,P_0^*)$ are mapped to target states $(T_1^*,P_1^*)$ drawn from the conditional grid, and the network parameters are optimized by minimizing the Kullback-Leibler divergence with respect to the corresponding Boltzmann distribution. Interestingly, in this setting, prior and reference distributions may coincide when the results are subsequently evaluated at $(T_0^*,P_0^*)$.

\subsection{Efficiency}
Alongside analyzing the loss function, the Wasserstein distance is computed at each epoch to assess the alignment between the generated and prior energy distributions. It is defined as

\begin{equation}
W_1(p'_A,p_B) = \int_{-\infty}^{+\infty} |F'_{A}(E) - F_{B}(E)| dE,
\label{eq:wasserstein} 
\end{equation}
where $F'_{A}(E)$ and $F_{B}(E)$ are the cumulative distribution functions of the generated and prior energy distributions, respectively. This distance measures the discrepancy between the distributions, with a smaller value indicating better alignment.

This integral formulation offers an intuitive metric for quantifying how much the generated energy landscape deviates from the prior distribution. However, since the energy scale can vary across different thermodynamic states, it is useful to introduce a relative Wasserstein distance that normalizes these variations. This is accomplished by scaling eq.~\ref{eq:wasserstein} with the range of energy values obtaining the relative Wasserstein distance:

\begin{equation}
    W_{rel} = \frac{W_1(p'_A,p_B)}{E_{max}-E_{min}}
    \label{eq:wasserstein_rel} 
\end{equation}
This normalization ensures that the Wasserstein distance remains meaningful and comparable across thermodynamic states, avoiding artificial inflation due to differences in energy scales. We then defined an efficiency metric as \textit{Efficiency} $= 1 - W_{rel}$. This quantity formally spans the interval $[1,-\infty)$, since Wasserstein distances can exceed the reference range. In practice, we interpret efficiency values below zero as indicating distributions whose separation exceeds the acceptable energy range, and are therefore considered poor matches. We employ this practical criterion, considering distributions with relative distances exceeding the reference threshold as non viable.

A key advantage of the Wasserstein distance is its connection to distributional convergence, making it a reliable indicator of training progress. As the model improves, the distance between the generated and true energy distributions decreases, reflecting the network's ability to learn the underlying energy landscape.
Furthermore, some pioneering work has explored using the Wasserstein distance as a training objective~\cite{frogner2015}, and investigating this approach, either alone or in combination with the KL divergence, represents a promising direction for future work.

To facilitate comparison with existing metrics, we also report the Effective Sample Size (ESS)~\cite{kish1995}, which estimates the number of statistically independent samples obtained from a weighted ensemble, defined as:
\begin{equation}
    \mathrm{ESS} = \frac{\left(\sum_{i=1}^{N} \omega_i \right)^2}{\sum_{i=1}^{N} \omega_i^2}.
\end{equation}
Where $\omega_i$ denote the weights associated with individual configurations. As we will show in the following, while the ESS provides useful information, it tends to decrease more abruptly and step-wise, and does not track the loss function as closely, highlighting the Wasserstein distance as a more geometry-aware, distribution-level measure of sampling efficiency.

\section{Liquid-Gas coexistence}

\begin{figure}[h!]
    \centering
    \includegraphics[width=1\linewidth]{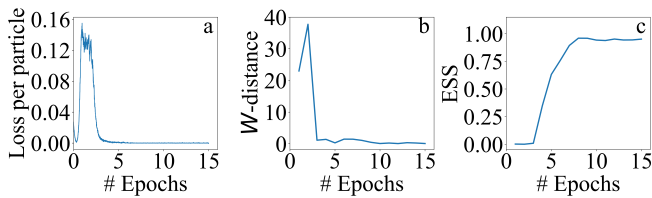}
    \caption{(a) Loss function, (b)  Wasserstein distance, and (c) Effective Sample Size as a function of epochs for the liquid state point training.}
    \label{fig:loss_wass_liquid}
\end{figure}

To benchmark and optimize the method, we initially evaluate its performance deep in the liquid phase. This enables systematic assessment of the model’s ability to generate physically realistic configurations both \textit{at} the training points and \textit{in their local vicinity}. Crucially, we investigate the method’s extrapolation range – i.e., how far from the reference state it can reliably sample configurations while maintaining physical fidelity.

\begin{figure}[h!]
    \centering
    \includegraphics[width=1\linewidth]{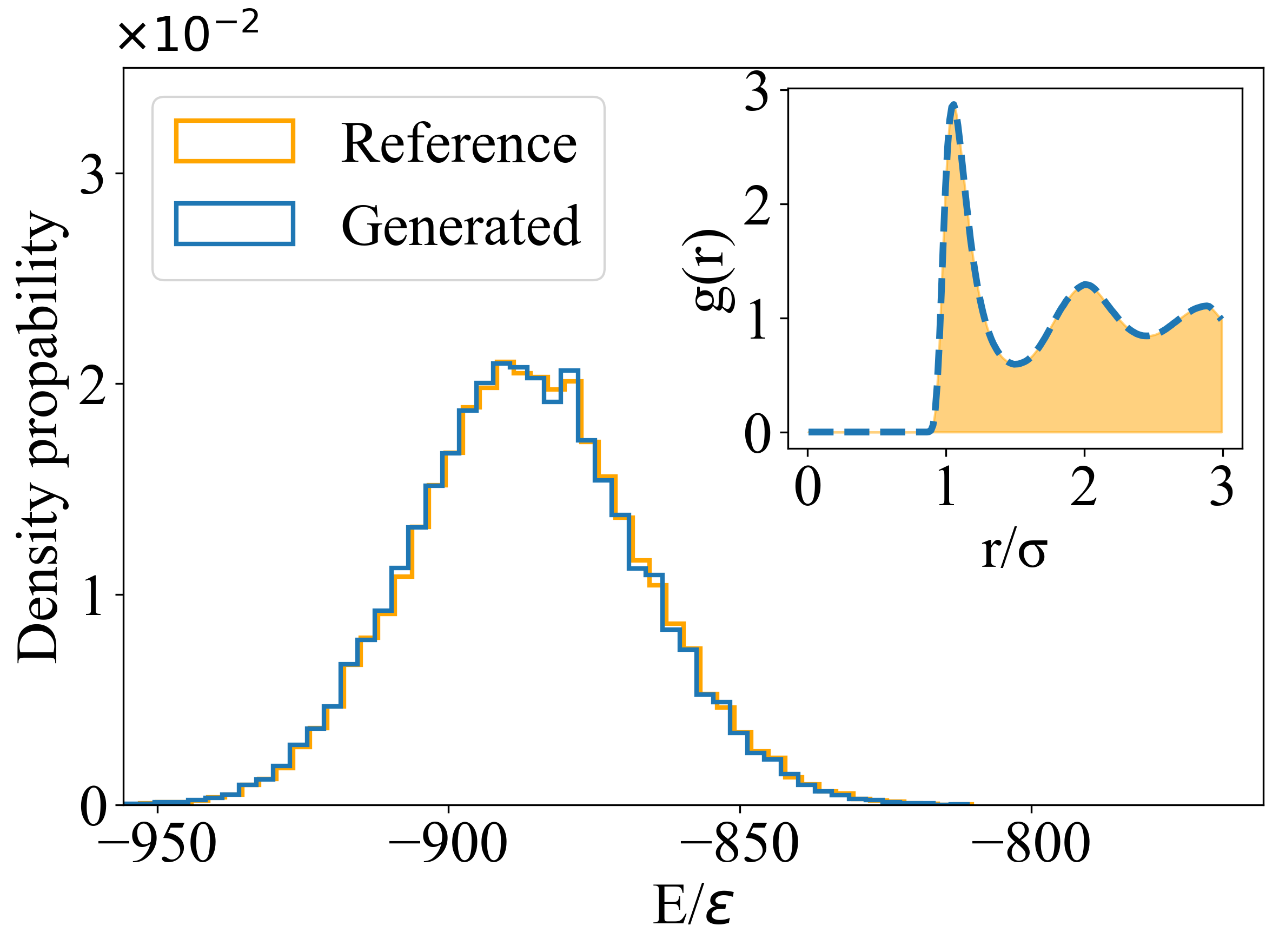}
    \caption{Energy density probability both for reference and generated samples. Inset: corresponding radial distribution function $g(r)$, reference results are shown as a shaded area.}
    \label{fig:ene_rdf_liquid}
\end{figure}

The training in the liquid state is carried out at $T^*=1.3$ and $P^*=6.5$, using a conditional grid of a $80$ by $80$ mesh over $T^*=\left[0.6,1.6\right]$ and $P^*=\left[4,20\right]$.  In this case, training progresses smoothly, with both the loss function and the Wasserstein distance approaching zero within just four epochs, while the ESS converges more slowly, reaching a value near one only after eight epochs, as illustrated in Fig.~\ref{fig:loss_wass_liquid}. Despite this rapid convergence, we extended the training to 15 epochs to further refine the model. During this process, we also monitored the potential energy distribution and the radial distribution function by generating samples at the training point. These were compared with reference results from standard simulations. We observed progressive convergence after epoch four. The final comparison, shown in Fig.\ref{fig:ene_rdf_liquid}, demonstrates a strong agreement between the generated and reference distributions for both energy and $g(r)$.

\begin{figure}[h!]
    \centering
    \includegraphics[width=1\linewidth]{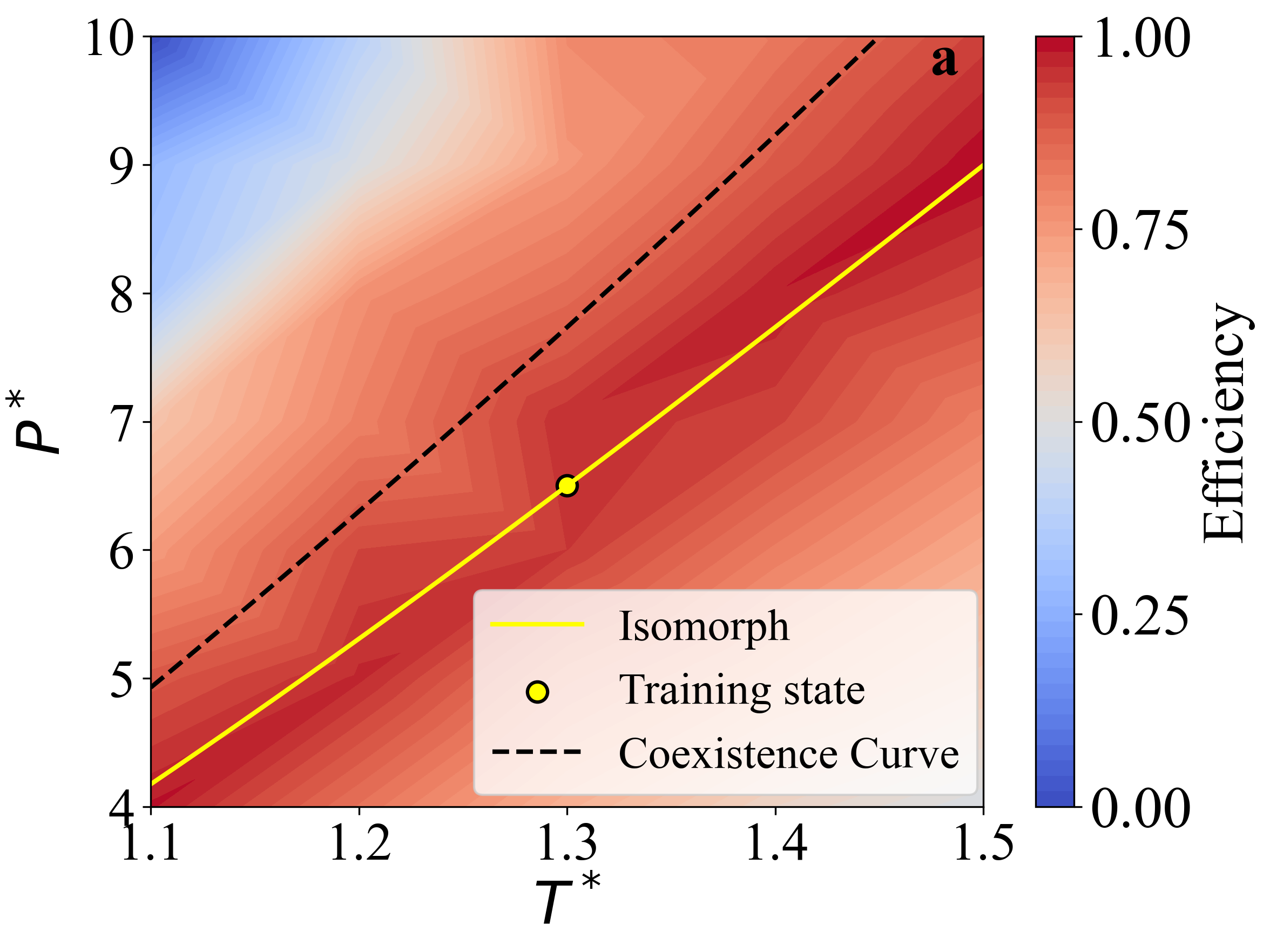}
    \includegraphics[width=1\linewidth]{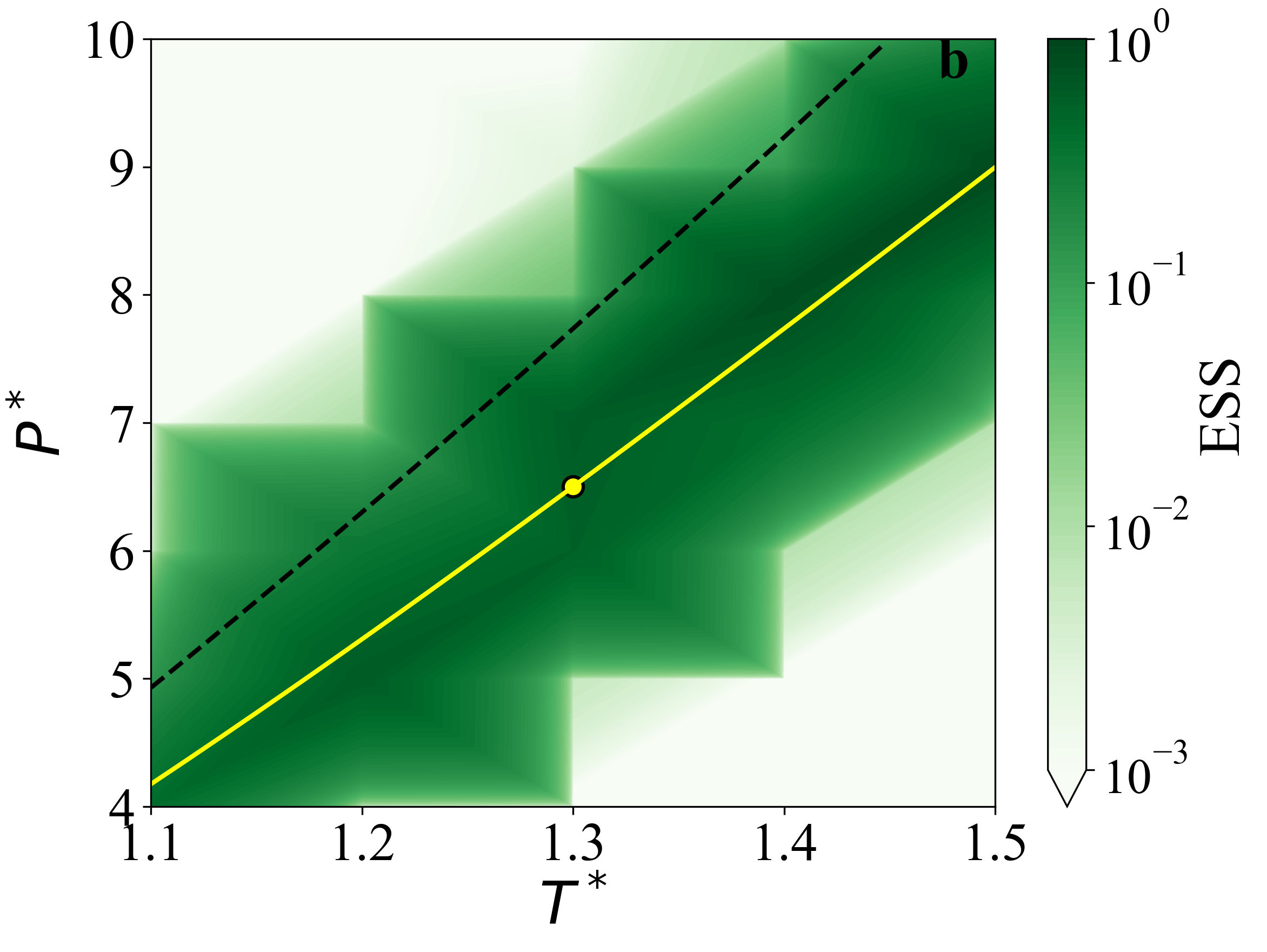}

    \caption{Efficiency maps of the liquid training as relative Wasserstein distances (a) and Effective Sample Size (b) between reference and generated configurations. The yellow point represent the training state point. As a conditional grid for efficiency evaluation we used a $5$ by $7$ mesh over $T^*=\left[1.1,1.5\right]$ and $P^*=\left[4,10\right]$. The black dashed curve, representing the liquid-solid coexistence, is adapted from Ref.~\cite{schebek2024,ccbylicense}. The yellow full curve represent the system isomorph.}
    \label{fig:map_liquid}
\end{figure}

After completion of the training, we evaluated the model’s generative performance across a broad range of temperatures and pressures to effectively sample the entire phase diagram. The {\it Efficiency} and the ESS metrics was evaluated, and the corresponding results are shown in Fig.~\ref{fig:map_liquid}.
As shown, the model achieves good performance across most of the investigated region of the phase diagram. Notably, the region of high efficiency appears to trace the liquid-solid coexistence line (dashed curve in Fig.~\ref{fig:map_liquid}), extending well into the metastable liquid regime. However, in the high-pressure, low-temperature region, where the crystalline phase is the thermodynamically stable and dominates the statistical ensemble, the sampling efficiency deteriorates rapidly. This trend hints at a connection between BG efficiency and underlying thermodynamic behavior, as performance begins to decline only after a fixed amount of supercooling.
A closely related physical interpretation was recently discussed in Ref.~\cite{coretti2025}, where normalizing flows were used to map configurations between WCA and Lennard–Jones liquids. In that work, high sampling efficiency was observed along thermodynamic paths for which the liquid structure is approximately preserved, and it was argued that in this regime the flow mainly learns an effective global transformation, while learning genuine many-body correlations becomes significantly more challenging when prior and target structures differ. Our results are fully consistent with this picture, where the efficiency of the Boltzmann Generator were highest along specific thermodynamic paths.

Following the concept of isomorphs~\cite{bailey2008, schroder2011}, which are exact for inverse-power-law potentials and provide an approximate description for Lennard-Jones systems, we compare the region of high sampling efficiency in the liquid regime with the corresponding isomorph obtained from standard isomorph tracing. For the liquid training point, the correlation coefficient between potential energy and virial is moderate ($R \simeq 0.5$), indicating that the system is outside the strictly strongly correlating regime where isomorph invariance is expected to be quantitatively accurate. Nevertheless, the isomorph follows the region of high sampling efficiency more closely than the liquid-solid coexistence line, as shown by the yellow solid curve in Fig.~\ref{fig:map_liquid}. This suggests that, even when strong virial-energy correlations are absent, isomorph-based thermodynamic paths provide a useful reference for rationalizing the regions in which the flow predominantly captures global structural transformations, as also noted in in Ref.~\cite{coretti2025}.

\section{Sampling the Critical point}

\begin{figure}[h!]
    \centering
    \includegraphics[width=1\linewidth]{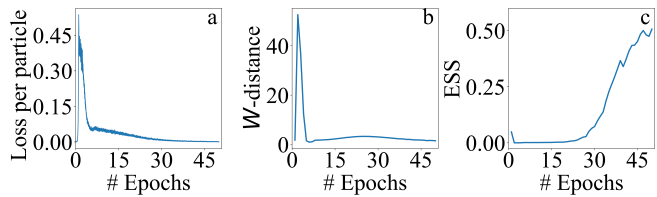}
    \caption{(a) Loss function, (b) Wasserstein distance, and (c) ESS as a function of epochs for the critical state point training.}
    \label{fig:loss_wass_critico}
\end{figure}

We now investigate the neural network’s ability to learn and generalize in the vicinity of the critical point. To estimate the critical point for our system, configurations were generated using isothermal-isobaric Monte Carlo simulations over the range $T^* \in [1.060, 1.180]$. The $P(\rho)$ equation of state was obtained and interpolated with a third-degree polynomial to identify the inflection points. The critical point, defined as the state with a single inflection point, was determined to be at $T^*_c = 1.1364$ and $P^*_c = 0.1163$.

We perform two types of training: in the first, the model is trained at the critical point and then used to generate configurations in its vicinity, allowing us to assess whether the network can learn the critical behavior directly. In the second, the model is trained near, but not at, the critical point and then used to generate critical configurations, providing insight into the network's ability to extrapolate to critical states it has not seen during training.

We start by conducting the training with a prior at the critical point and a conditional grid of a $500$ by $500$ mesh over $T^*=\left[0.6,1.6\right]$ and $P^*=\left[0,4\right]$. As illustrated in Fig.~\ref{fig:loss_wass_critico}, both the loss function and the Wasserstein distance converge more slowly compared to the liquid case. The ESS appears to be even more sensitive, reaching full convergence only after approximately $50$ epochs. We also carried out preliminary tests using a coarser $80$ by $80$ grid, which resulted in poorer training performance.

\begin{figure}[h!]
    \centering
    \includegraphics[width=1.\linewidth]{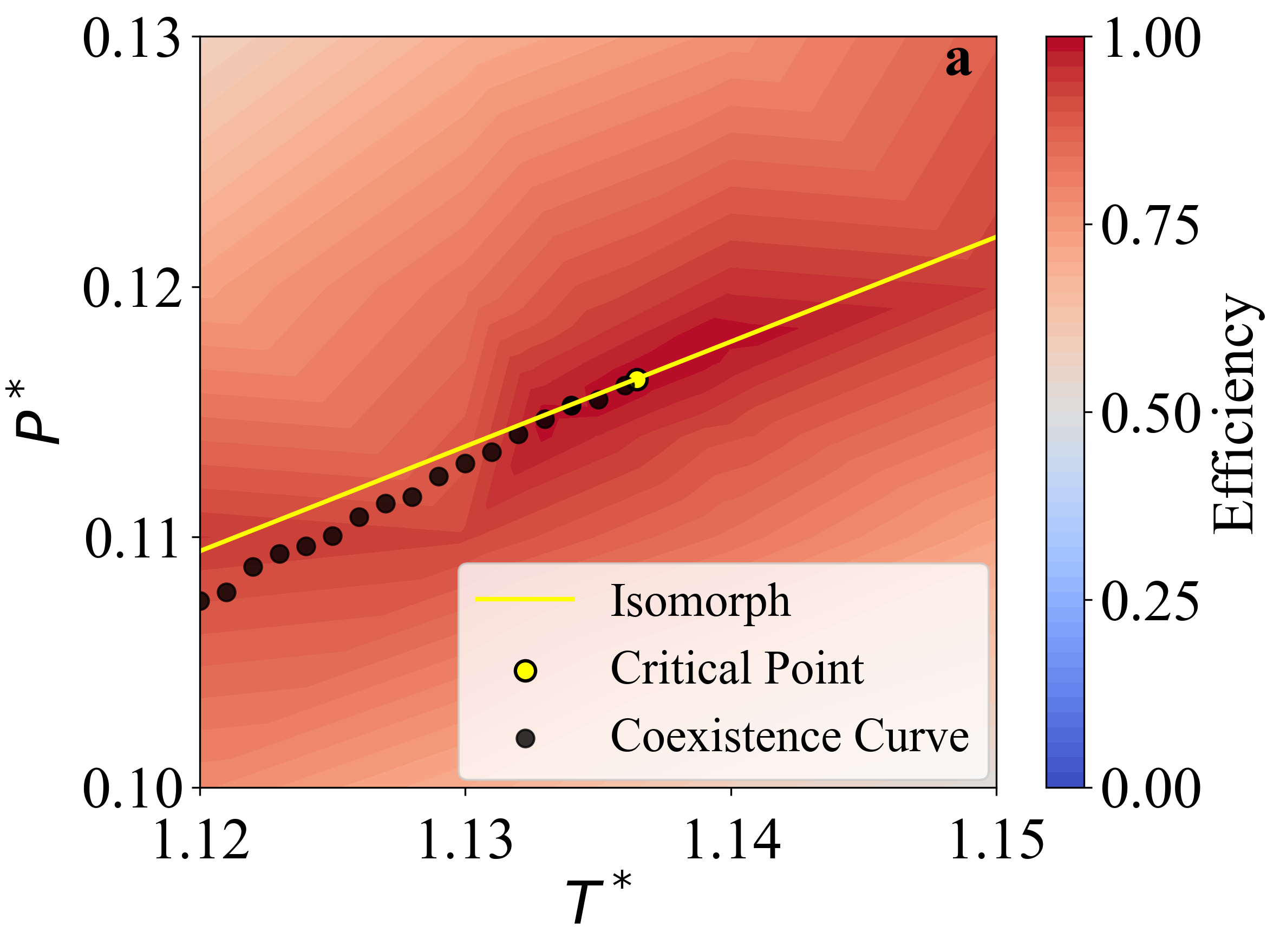}
\includegraphics[width=1.\linewidth]{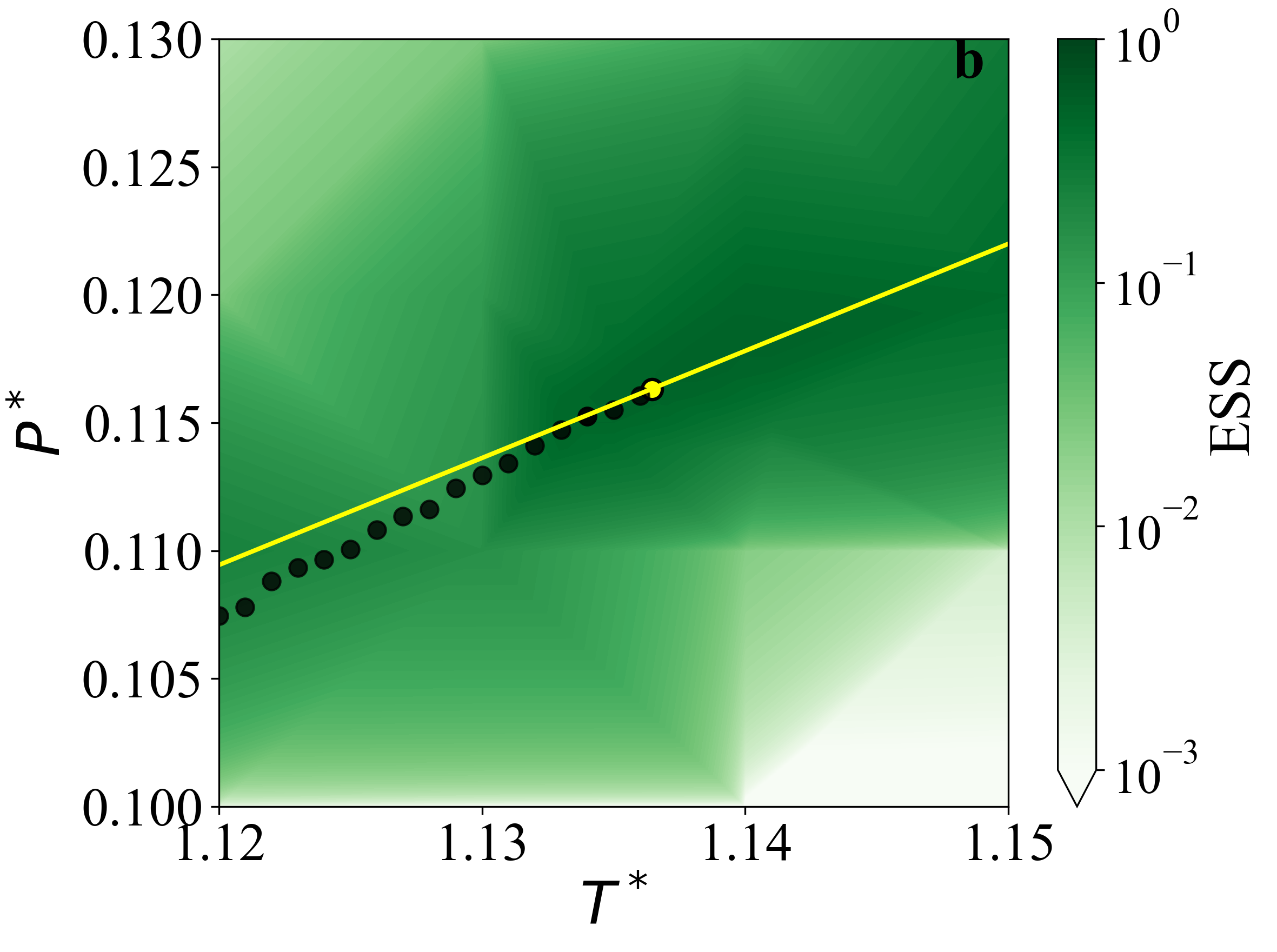}
\caption{(a) Efficiency map for training at the critical point, shown in terms of relative Wasserstein distances between the reference and generated configurations. The yellow point denotes the training state corresponding to the critical point . For the conditional evaluation of the efficiency, we combine a $4 \times 4$ grid spanning $T^* \in [1.12,1.15]$ and $P^* \in [0.10,0.13]$ with a $5 \times 5$ grid spanning $T^* \in [1.133,1.141]$ and $P^* \in [0.114,0.118]$. Grey points indicate the coexistence line as determined from the Maxwell construction, while the yellow line denotes the system isomorph. (b) An analogous representation of the ESS map.}
\label{fig:map_critical}
\end{figure}

\begin{figure}[h!]
    \centering
    \includegraphics[width=1\linewidth]{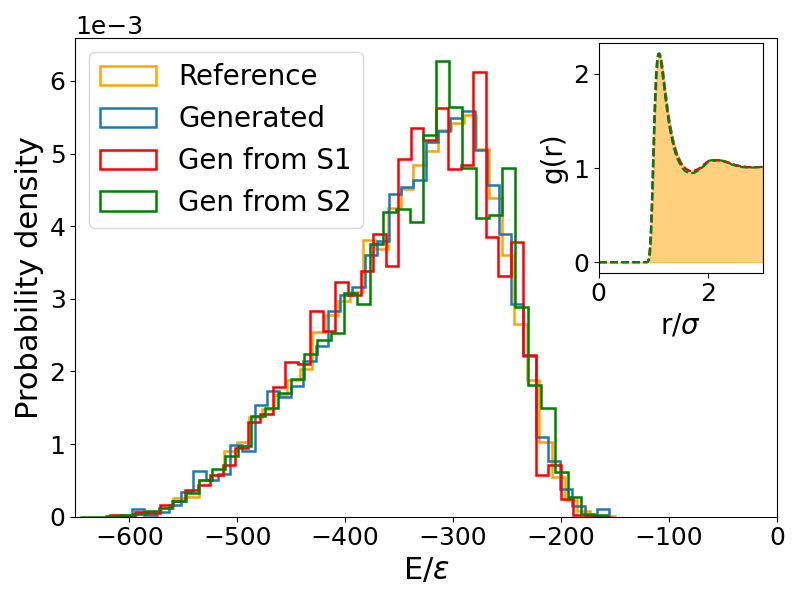}
    \caption{Probability distributions of the energy density for both reference and generated samples are shown at the critical point, as well as at states $S_1$ and $S_2$, represented by orange, blue, red, and green curves, respectively. The inset displays the corresponding radial distribution functions $g(r)$ for each case and the reference results are shown as a shaded area.}
    \label{fig:ene_rdf_critico}
\end{figure}

We now examine the efficiency map in the $P$–$T$ plane near the critical point. As shown in Fig.~\ref{fig:ene_rdf_critico}, the network exhibits high sampling efficiency within a small region surrounding the critical point. By overlaying the liquid-gas coexistence line (determined from Maxwell constructions on the isotherms) onto the efficiency map, we note that the performance decline of the network {\it Efficiency} roughly follows the same trend. This shows once again that the ability of the network to generate representative configurations is correlated with the underlying phase behavior, as previously noted for the liquid configurations near the melting line (Fig.~\ref{fig:map_liquid}). Also near the critical point, the efficiency remains high changing temperature or pressure, allowing for generation in sub- and super-critical regions. As observed for the liquid, the ESS here also follows the trend indicated by the Wasserstein distance, although its decrease is smoother and the values remain comparatively higher and above $10^{-3}$.

Similarly, an analysis based on isomorph theory can be carried out near the liquid-gas critical point. In this regime, the virial-energy correlation increases to $R \simeq 0.7$, but still remains below values typically associated with strongly correlating liquids. Despite the presence of pronounced critical fluctuations, the isomorph continues to provide a reasonable description of the region of enhanced sampling efficiency, as indicated by the yellow solid curve in Fig.~\ref{fig:map_critical}. Close to criticality, however, the numerical determination of isomorphs becomes increasingly noisy due to large fluctuations. Overall, these observations indicate that approximate isomorphic invariance remains relevant for understanding the performance of the flow across different thermodynamic regimes.~\cite{coretti2025}

We now investigate whether BGs can generate configurations at the critical point when using priors located away from criticality.
To this aim, we perform the training at four ($T^*$ , $P^*$)  thermodynamic states, $(S_1=1.139,0.117)$, $(S_2=1.12,0.11)$, $(S_3=1.15,0.11)$, and $(S_4=1.15,0.10)$, positioned progressively farther from the critical point. Once trained, we use the network to generate configurations at the critical point. Training at $S_1$ yields energy distributions and pair–correlation functions that approximate the critical state with reasonable fidelity. Training at $S_2$ produces an energy histogram with a bimodal structure, characteristic of subcritical regimes. After reweighting, both cases show good agreement with the target distribution, as illustrated in Fig.~\ref{fig:ene_rdf_critico}, which exhibits pronounced tails toward lower energies, consistent with the enhanced fluctuations expected near criticality in finite systems. In contrast, training at $S_3$ and $S_4$ results in networks that fail to generate meaningful configurations at the critical point, producing energy distributions that are extremely noisy and statistically insignificant. For clarity, these are not shown in the figure. 
These results emphasize the challenges in extrapolating across phase boundaries, especially near the critical point. While direct training performs well overall, we find that enhanced generalization in these regions requires targeted retraining or increased sampling density in thermodynamic space.

\begin{figure}[h!]
    \centering
    \includegraphics[width=1.\linewidth]{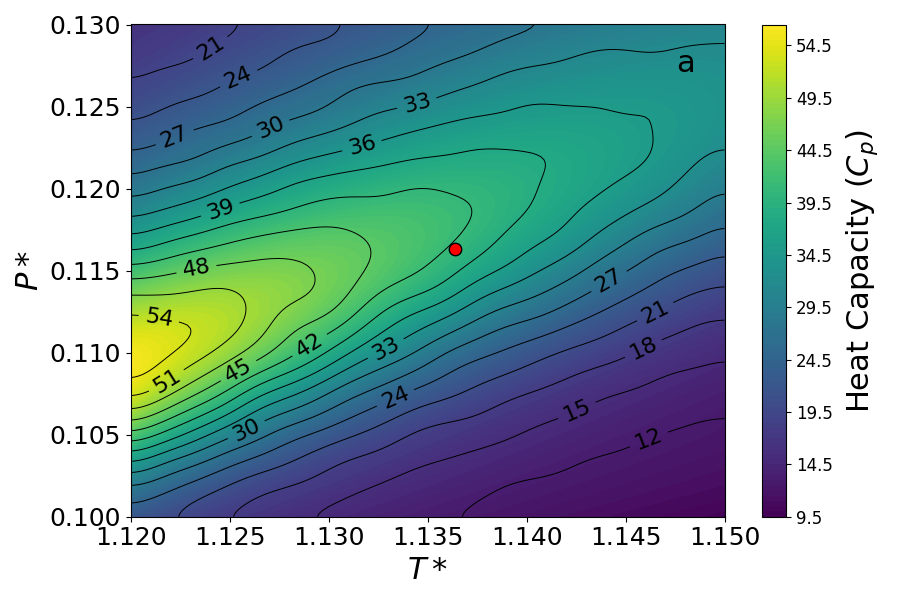}
    \includegraphics[width=1.\linewidth]{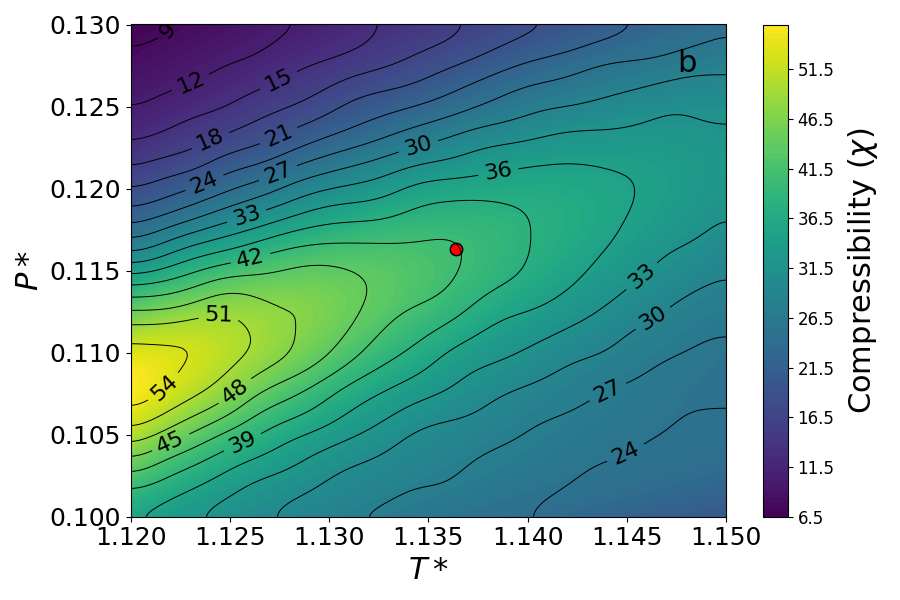}
\caption{Heat capacity (top) and compressibility (bottom) in the vicinity of the critical point (red point). Contour lines correspond to MC results, while the color maps are derived from generated configurations.}
\label{fig:map_cp_kappa}
\end{figure}

Having analyzed the energy behavior so far, it is both more insightful and statistically challenging to investigate the associated fluctuations. We compute both the heat capacity at constant pressure and the compressibility, respectively given by
\begin{equation}
    C_p = \frac{\langle H^2 \rangle - \langle H \rangle^2}{k_\mathrm{B} T^2},\hspace{0.2cm}
    \kappa_T = \frac{\langle (\Delta V)^2 \rangle}{k_B T \langle V \rangle}
\end{equation}
where \(H\) denotes the enthalpy, \(T\) the temperature, \(V\) the volume and \(k_\mathrm{B}\) the Boltzmann constant. The quantity \(C_p\) and $\kappa_T$ are evaluated using Monte Carlo (MC) sampling with 20,000 configurations distributed over a \(20 \times 20\) grid within the interval shown in Fig.~\ref{fig:map_critical}. An equivalent number of samples is then generated using the network trained at the critical point discussed in the previous section, then \(C_p\) and $\kappa_T$  are computed in the same manner. The comparison in Fig.~\ref{fig:map_cp_kappa} shows the values obtained from the generated data as a color map and those from MC as contour lines. As observed, both \(C_p\) and $\kappa_T$ reach a pronounced maximum in the vicinity of the critical point and along the extension of the so-called \emph{Widom line}, consistent with the expected behavior of response functions near critical phenomena. We also observe that the $C_p$ maximum line does not precisely trace the backbone of critical fluctuations, a behavior that can be understood in terms of finite-size scaling~\cite{cardy1988}, due to systematic shifts related to the slope of the coexistence line~\cite{stanley2014}, and intrinsic properties of the model~\cite{brazhkin2011}.
Overall, both the heat capacity and compressibility computed from MC simulations are well reproduced by the generated data, in both magnitude and shape. This highlights the BG model’s ability to accurately capture equilibrium fluctuations near critical points.

\section{Conclusions}
Recent advances in AI-based generative modeling open the door to new opportunities for simulating and understanding complex liquids and soft matter systems. While substantial progress has been made in fields such as protein science,~\cite{noe2019,lewis2025,murtada2025} the application of these methods to high-dimensional, disordered systems remains limited, with relatively few attempts~\cite{jung2024,jung2025} and some notable challenges.~\cite{ciarella2023}

In this work, we demonstrated the application of Boltzmann Generators to model and sample configurations at the liquid-gas critical point of a Lennard-Jones fluid. Our approach successfully generated critical states and reproduced equilibrium properties in regions that are typically hindered by critical slowing down. By conditioning on thermodynamic variables, BGs effectively generalized across neighboring states and provided a compelling alternative to traditional simulation methods.

We demonstrated substantial computational advantages: while traditional Monte Carlo sampling of 20,000 configurations requires around 20 hours on a single CPU core, BGs trained on the liquid state complete in under three hours on a modern GPU and can generate new configurations in under 10 minutes. These results suggest that generative models can enable rapid exploration of phase space, potentially unlocking studies at longer timescales. At the same time, we identified clear limitations in maintaining robustness when extrapolating far from the critical point, thereby highlighting opportunities for methodological improvement.

Future work may focus on hybrid strategies combining BGs with traditional simulations, adaptive training schemes~\cite{gabrie2022}, and extensions to glass transitions, nucleation, and nonequilibrium phenomena. Further development could improve scalability, robustness, and transferability across different systems. More broadly, our results highlight the promise of generative neural networks to complement and accelerate classical simulation methods, offering a pathway toward more efficient and flexible exploration of complex phase behavior.

\section{Acknowledgements}
We acknowledge the CINECA award under the ISCRA initiative, for the availability of high performance computing resources and support.

\nocite{*}
\bibliography{bibliography2}

\end{document}